# Influences of Uncertainties in Thermodynamic Models on Pareto-optimized Dividing Wall Columns for Ideal Mixtures

Lea Trescher[1], David Mogalle[2], Patrick Otto Ludl[2], Tobias Seidel[2], Michael Bortz[2], Thomas Grützner[1]

[1]Ulm University, Institute of Chemical Engineering, Laboratory of Thermal Process Engineering, Ulm, Germany.

[2]Fraunhofer Institute of Industrial Mathematics (ITWM), Kaiserslautern, Germany.

* Correspondence: lea.trescher@uni-ulm.de

Abstract

This article examines the effect of individual and combined uncertainties in thermodynamic models on the performance of simulated, steady-state Pareto-optimized Dividing Wall Columns. It is a follow-up of the previous work analogously treating deviations in process variables. Such deviations and uncertainties that may even be unknown during the design process can significantly influence the separation result. However, other than process variables, uncertainties in thermodynamics are usually not systematically considered during design. For the first time, the effects of uncertain thermodynamic properties on Pareto-optimized DWCs with different numbers of stages and for different mixtures are presented and compared qualitatively and quantitatively. Depending on the number of stages and mixture characteristics, particularly critical properties are identified. On the one hand, this provides information on aspects requiring special attention prior to design, and on the other hand, it also indicates in which section of the DWC a stage supplement might be most beneficial.

## Keywords

## Highlights

- Models related to the vapor-liquid equilibrium have the decisive influence.
- Caloric properties are unsignificant in the investigated range of uncertainty.
- Affected VLE or internal liquid to vapor ratio leads to stage-dependent losses.
- Vapor pressure of medium boiler is decisive independent of mixture or stages.
- VLE of binary split with highest minimum vapor demand is additionally crucial.

# 1 Introduction

Due to the continued importance of distillation as separation process in the chemical and process industry and the associated high energy requirements, the application of intensified energy-saving processes is more relevant than ever. Dividing Wall Columns (DWCs) for the separation of ternary mixtures have been researched and industrially established for decades, their further development to quaternary mixtures as multiple Dividing Wall Columns (mDWCs) is limited to the pilot scale so far [1]. For a detailed introduction to DWCs the reader is referred to the literature [2 - 8], a short summary can also be found in the previous article of the authors [9]. The saving potentials compared to conventional sequences are up to 30 % for DWCs and 50 % for mDWCs in both CAPEX and OPEX - provided that the column design and operation are at the optimum.

The design of a distillation column for a specific separation task is in theory possible between the two extreme points of minimum number of theoretical stages (at infinite reflux) and minimal reflux (with an infinite number of stages). Both have no practical relevance due to a lack of product flows / excessive energy input or investment costs, respectively. Optimized solutions, i.e., columns with the lowest possible energy input for a given number of stages, lie between the two extreme values on the so-called $NQ$ curves. These represent one possibility of a Pareto front, i.e., a set of design points, where no objective can be improved further without compromising another. The design decision is then usually a trade-off between the two objectives theoretical stages $N$ (correlating with CAPEX) and reboiler duty $Q$ or reflux ratio $R$ (correlating with OPEX). Rules of thumb such as a total number of theoretical stages $N$ of 2 to 2.5 · $N_{min}$ or a reflux ratio of approx. 1.05 to 1.5 · $R_{min}$ [10 - 13] exist for the range of interest. However, the design is not selected directly on the Pareto front; instead, safety margins are applied for both the theoretical stages and the energy input in order to cover possible uncertainties and to ensure that the required specifications are met. For such uncertainties, two main sources of deviations from the design case of a plant exist: process variables (such as the exact feed flow and composition or the pressure (drop)) can deviate permanently from the design case or fluctuate, and the thermodynamic models used for design can be subject to errors and uncertainties. The latter can be caused either by systematic or random measurement deviations, but also by model errors that persist even with the greatest effort for good data and may remain unknown; no distinction is made in the following. Deviations in process variables are recognized and usually covered by sensitivity studies and/or safety margins. Accuracies of thermodynamic data or fitted property models, on the other hand, are also treated in literature since the early 1980s. Larsen [14] discusses the evaluation and improvement of available experimental data in 1986 and already emphasizes that the effects of data errors can be "insignificant or enormous" depending on the process and mixture. Activity coefficients are mentioned, where an uncertainty of around 10 % in the case of a high separation factor and low stage requirement can mean a negligible error in the resulting costs, but in the case of a separation factor below 1.1 can mean a factor of 2 for required stages or resulting costs. Nevertheless, the systematic consideration of these uncertainties is not a common part of process design practice: On the one hand, especially mixture properties pose challenges and contain residual uncertainties even with high-quality data [15]. On the other hand, it is usually not possible to simply adjust properties - especially such as activity coefficients - in conventional process simulators. Instead, an adapted parameter regression would be necessary for each extent of uncertainty. Mathias [15 - 17] presents a method for quantifying the uncertainty in phase equilibria and a method based on the Margules formula for perturbing the activity coefficient, whereby the percentage deviation is higher the further the nominal value deviates from 1 (i.e., from ideality). In three application examples, a strong dependence of the effects on the mixture was determined; with significantly higher inaccuracy in the thermodynamic model, the design inaccuracies of a dehexanizer column were considerably lower than those of a propane-propylene (the significantly closer boiling mixture) splitter.

Furthermore, the equilibrium constants were found to have a higher influence on utility demand than on the flowsheet structure. In [16], different pairings of activity coefficient uncertainty were examined; for deviations in the same direction (i.e., both positive or negative), only little influence was observed. Burger et al. [18] propose a similar perturbation of the fugacity coefficient for multicomponent mixtures and present examples of the effect of uncertain pure vapor pressures and a VLE model. In the range of vapor pressures investigated, an approximate doubling of both minimum reflux and minimum number of stages was found for a distillation column for the separation of Cumene + p-Diisopropylbenzene at 67 mbar between best and worst case - however, uncertainties of up to approx. 50 % of the pure substance vapor pressures were investigated. In another example, the effect of ± 10 % in pressure on the gas solubility of carbon dioxide in methanol is investigated. The closer to the minimum solvent requirement, the greater the influence on the stage requirement; from approx. 2.5 times the minimum solvent requirement, there is no longer an influence.

All previous statements on design, optimization and presence of uncertainties apply to DWCs as well as to simple columns. However, in case of DWCs, design and underlying optimizations are complicated on the one hand by additional degrees of freedom: Number of stages in four more sections, an additional product flow as well as vapor and liquid split. On the other hand, these internal splits and especially the vapor split *VS*, represent a particularity when balancing a compromise between proximity to the optimum and a sufficient safety margin: The vapor split usually occurs due to the flow resistance in the parallel column strings next to the partition wall, i.e., it is fixed after construction and can usually neither be measured nor adjusted. The fulfillment of the separation task, though, depends eminently on the appropriate vapor split or at least a suitable combination of vapor and liquid split *LS*, and the closer the column is to the energetic optimum, the narrower the feasible operating window (i.e., possible combinations of *VS* and *LS*) [19]. An unfavorable distribution of theoretical stages, depending on the specific separation case, can also lead to a reduced or shifted operation window [20]. The same accounts for deviations from the nominal design, changes in the position and extend of the feasible operation window can results. This shift can lead to a situation, where the actual *VS* (fixed by construction) lies outside the feasible range. I.e., the required separation is no longer possible, even with increased energy input. The design of a DWC must therefore ensure that within the scope of expected uncertainties, it is still possible to achieve a feasible *LS*/*VS* combination by adapting the *LS*. Such shifts in the operation window were presented, for example, for deviating feed compositions by Ge et al. [21]. To the authors' knowledge, there are no systematic overviews of the relationship between mixture properties, resulting optimized structures of DWCs and their influence on the sensitivity to different possible deviations from the design case. For uncertainties in thermodynamic models, such investigations for DWCs are not known to the authors.

This work is the second part of a research that focuses on systematically investigating the relationships between the proximity of DWC configurations to the energetic optimum, their sensitivity to deviations from the nominal design case and the remaining flexibility in operation to fulfill the separation task. This is intended to provide a fundamental understanding of which factors lead to severely restricted flexibility, i.e., in which cases a high additional energy input is required or the separation task can no longer be fulfilled at all - and, above all, how this can be prevented. In other words, the overall leading question is: How must a DWC generally be designed in relation to the separation task in order to be as close as possible to the energetic optimum, but also with low CAPEX, while at the same time having a large feasible operation window that is stable even under deviations?

In the previous work [9], the same investigations were carried out with deviating process variables as are presented here for thermodynamic models. However, this closely related paper contains some preliminary work referred to here: The mixtures that are also used here were characterized and the *NQ*-curves obtained by multi-criteria optimization were shown. These are limited to mixtures with close-to-ideal behavior, i.e., the activity coefficients of the liquid mixture are approx. constant ≈ 1. Real mixtures pose additional requirements, e.g., with regard to the initialization of the multi-objective optimization, which needs more discussion than possible here. Real mixtures will therefore be part of a subsequent work. Four DWCs with different total stage numbers were examined for each mixture and evaluated with regard to their stage distribution, which was found to be unexpected in some cases. When applying deviations from the nominal case, it was found that if the ratio of internal flows is affected (i.e., liquid and vapor split), the purity losses are stage-dependent - the higher the total number of stages, the greater the loss for at least two of the three products and that in this case, the side product is mixture-independently most sensitive. Effects of combined deviations ("worst-case scenarios") were found to be approximately an addition of the individual effects. All simulations, both in [9] and in this present work, are limited to steady-state simulations. Verifications using dynamic models are part of ongoing work.

In this subsequent work, the effects of uncertainties in thermodynamic models from the nominal case on the same steady-state, Pareto-optimal columns are investigated. It is organized as follows: Section 2 gives again a brief description of the mixtures and the four columns each, as well as an explanation of the uncertainties investigated and their application. Section 3 then presents the results, first for individual uncertainties and then for combined scenarios. It is shown that the influence of uncertain thermodynamic models depends also on the total number of stages of a Pareto-optimized DWC, if the vapor-liquid-equilibrium (VLE) or the ratio of internal flows is influenced. Again, the side-product is most sensitive in these cases. Properties directly related to the VLE were found to have a significantly larger effect than the ones in caloric properties. In addition, $V_{min}$ diagrams by Halvorsen and Skogestad [22], which have been intensively used and presented in the past [23], once again provide important information: the products that are part of the binary split with higher minimum energy requirements are those with the highest purity losses. At the same time, the precision of the associated binary VLE, especially in the range of low mid-boiler concentrations, is crucial for avoiding worst cases.

## 2  Methods

### 2.1  Separation Task and Nominal Design Case

The separation task and the boundary conditions that describe the nominal design case (in the following: nominal case) are listed below. The corresponding optimized column configurations (nominal columns) were calculated on this basis in the previous work [9].

- Product purities $x_i \geq 0{,}95$ mol/mol
- 3 kmol/h equimolar feed
- Feed is boiling liquid
- 1 bar operating pressure without pressure losses in the column
- Adiabatic column behavior
- Equilibrium stage model with MESH equations, model parameters used are retrieved from Aspen Plus® V11 (in the following Aspen) from databank APV110VLE-IG. Modelling includes:
    - Activity coefficients of all components in mixtures are calculated with the non-random two liquid model (NRTL).
    - Pure substance vapor pressures are calculated with the extended Antoine equation.
    - The enthalpies of vapor and vaporization are calculated using DIPPR model 801. The liquid enthalpy as the difference is not considered individually.

### 2.2  Investigated Mixtures

As already mentioned, the same mixtures with close-to-ideal behavior (i.e., activity coefficients in the liquid mixture $\gamma_i \approx \gamma_j \approx 1$) as in the previous work are used. The light boiler is designated as component A, the medium boiler as B and the heavy boiler as C. The mixtures including key data are presented in Tab. 1. The relative volatilities in each case are retrieved from Aspen for the feed mixture, the minimum stage requirements were calculated using Fenske equation. That this procedure provides sufficiently good results for ideal mixtures and that and how Fenske's equation can be used for the different pseudo-binary separations in a DWC has been shown and discussed in previous work [9, 20]. Both for the initialization of the multi-criteria optimization and for understanding mixture characteristics, the minimum vapor demand of the different binary separations as well as the optimum distribution of component B in case of a sharp AC split is important. This information is provided by $V_{min}$ diagrams, a shortcut calculation based on the assumption of infinite stages, pure products and constant relative volatilities. These diagrams were introduced by Halvorsen and Skogestad [22], detailed information, especially with regard to the significance for separations in DWCs, can also be found in [23]. The $V_{min}$ diagrams of the mixtures used were shown and discussed in the previous work [9]. Only the $V_{min}$ values for each sharp separation (i.e., the minimum required feed-related boil-up $V/F$) are listed here. It is important to note that for Systems 1 and 3 the BC split has the highest vapor demand and also a slightly higher minimum number of stages, while in System 2, the AB split requires most vapor and has by far the highest minimum number of stages. The nominal VLE (i.e., as calculated with the nominal NRTL model as defined in Section 2.1) of the binary sub-mixtures are shown in Fig. 1 as *T-x-y*-diagrams.

Table 1: Mixtures examined including relative volatilities of the feed mixture, minimum stage and vapor requirements (calculated with Fenske equation and $V_{min}$ diagrams, respectively) and boundary activity coefficients.

| System | Components A, B, C | Relative Volatilities $\alpha_{Feed}$ [AC;AB;BC] | Minimum Number of Stages $N_{min}$ AC, AB, BC | Minimum Vapor Requirement $(V/F)_{min}$ AC, AB, BC | Boundary Activity Coefficients $\gamma_{ij}^{\infty} / \gamma_{ji}^{\infty}$ | | |
|---|---|---|---|---|---|---|---|
| | | | | | AC | AB | BC |
| 1 | Benzene, Toluene, p-Xylene | [5.76;2.50;2.19] | 6, 7, 8 | 0.63, 0.82, 1.18 | 0.943/0.917 | 0.983/1.023 | 0.949/0.950 |
| 2 | Methanol, Ethanol, Butanol | [8.36;1.79;4.56] | 5, 11, 4 | 0.64, 1.23, 0.88 | 1.007/1.229 | 0.986/1.132 | 1.100/1.089 |
| 3 | Hexane, Heptane, Octane | [5.43;2.48;2.20] | 6, 7, 8 | 0.69, 1.01, 1.13 | 0.703/0.770 | 0.974/0.968 | 0.938/1.000 |

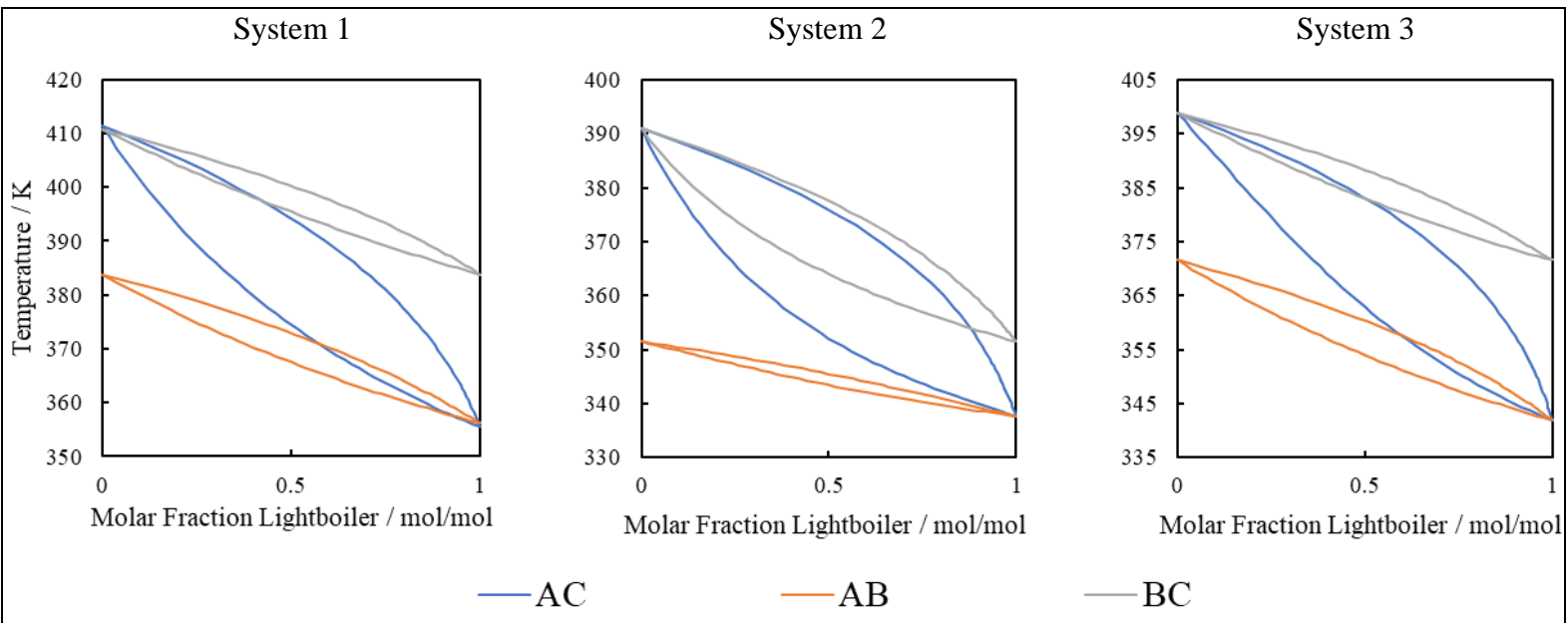

Figure 1: Binary *T-x-y*-diagrams of mixtures used. Data calculated in Aspen Plus using NRTL.

System 1 and 3 are clearly very similar with the boiling temperatures of System 1 being roughly 10 K higher. The AC separation is the easiest binary separation (in System 3 it is slightly non-ideal), AB and BC are practically identical. In System 2, however, the binary AC mixture is slightly, BC much wider boiling and the AB mixture is significantly closer boiling than in the other two mixtures – as already indicated in the minimum stage and vapor requirements. The close boiling points of A and B increase the similarity between AC and BC-VLE compared to the other two mixtures (see especially dew lines at high C content). When investigating the vapor demand of different splits at finite stages (Supporting Material 1 of the previous paper), a corresponding peculiarity was found: Although in all three mixtures, depending on the purity requirements, the AC split can have the highest vapor demand in some ranges (especially at very low numbers of stages), the optimal distribution of the mid boiler significantly changes only in System 2: Towards the minimum number of stages, the preferred AC split corresponds to an almost complete removal of B at the top. This also becomes clear in the Pareto-optimized columns - with regard to *VS* and *LS*, one point stands out with significantly increased split values, which is the column with only $1.2 \cdot N_{min}$ of System 2, see the following section.

## 2.3 Optimized Nominal Columns

For the above introduced separation task and each mixture, *NQ*-curves were calculated in the previous work [9]. From each curve, four Pareto-optimal column designs with different total stage numbers and accordingly different vapor demands were chosen, providing all specifications for the simulations (reboiler duty, product flows A and B, vapor and liquid split and stage allocations). These chosen columns are marked in Fig. 2 with the dashed rectangle. On the one hand, this covers the range typically considered for design; on the other hand, previous simulations have shown that simulations of columns even closer to the minimum number of stages lead to considerable convergence problems, while there are hardly any significant changes when moving towards an even higher number of stages. In the following, the columns with 25 and 26 stages respectively will be denoted to as columns with low, the ones with 56 and 60 stages as columns with high stage numbers.

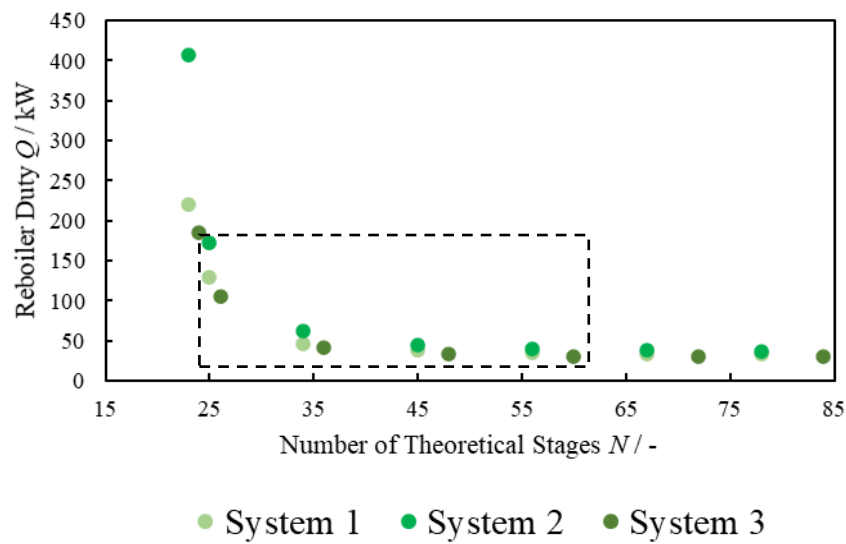

Figure 2: Pareto-curves for the separations of Systems 1 - 3 in products with 95 mol% purity in DWCs. Dashed rectangle: Pareto-optimal columns used in the following.

In order to understand the designations used in this work, Fig. 3 shows a scheme with the predominant splits assigned to each section. Products are in the following always designated A, B or C according to their main component. Tab. 2 summarizes the specifications of the simulated columns, a detailed description and discussion of the stage distribution to these different sections of the chosen DWCs in dependence on the total number of stages can be found in [9].

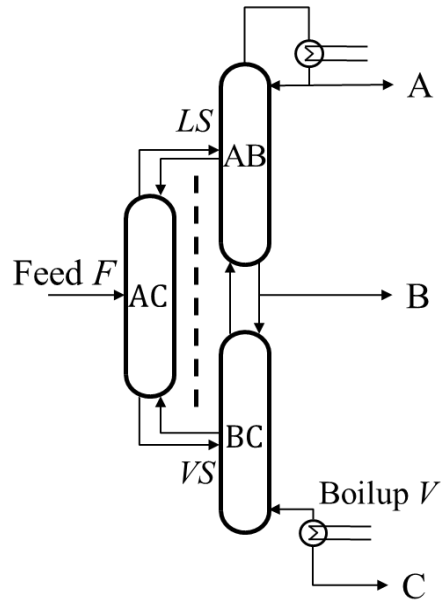

Figure 3: Scheme of DWC with binary splits assigned to the different sections. The dashed line represents the partition wall in this thermally coupled model.

Regarding the internal splits it should be noted that the values for the *VS* are generally slightly increasing with increasing stage numbers and in the range of 0.50 to 0.65. The values for the *LS* are slightly decreasing in the range of 0.45 to 0.30. An exception is the column with 25 stages (ca. $1.2 \cdot N_{min}$) of System 2: both splits are approx. 0.75 here, i.e., 75 % of vapor and liquid are required in the prefractionator. This also leads to a difference in the composition profiles of the columns. In Fig. 4, the nominal composition profiles are shown for the columns with low and high stage number for each of the three mixtures.

Table 2: Specifications of all optimized DWCs used for simulations.

| | Theoretical Stages $N$ / - | Reboiler Duty $Q$ / kW | AC Section | | AB Section | | BC Section | | Product A / mol/s | Product B / mol/s | $VS$ / - | $LS$ / - |
|---|---|---|---|---|---|---|---|---|---|---|---|---|
| | | | $N$ | $N_{Feed}$ | $N$ | $N_{Feed}$ | $N$ | $N_{Feed}$ | | | | |
| System 1 | 25 | 130.0 | 6 | 4 | 10 | 6 | 9 | 5 | 0.2896 | 0.2624 | 0.563 | 0.449 |
| | 34 | 46.2 | 9 | 5 | 11 | 5 | 14 | 6 | 0.2856 | 0.2624 | 0.564 | 0.356 |
| | 45 | 37.7 | 12 | 6 | 12 | 5 | 21 | 7 | 0.2872 | 0.2624 | 0.594 | 0.334 |
| | 56 | 35.5 | 14 | 8 | 12 | 6 | 30 | 9 | 0.2861 | 0.2623 | 0.612 | 0.331 |
| System 2 | 25 | 171.9 | 5 | 2 | 14 | 10 | 6 | 3 | 0.2814 | 0.2625 | 0.739 | 0.753 |
| | 34 | 62.0 | 9 | 3 | 19 | 10 | 6 | 4 | 0.2807 | 0.2626 | 0.527 | 0.323 |
| | 45 | 44.3 | 15 | 3 | 23 | 11 | 7 | 4 | 0.2814 | 0.2625 | 0.640 | 0.403 |
| | 56 | 39.5 | 19 | 4 | 29 | 14 | 8 | 4 | 0.2805 | 0.2624 | 0.649 | 0.367 |
| System 3 | 26 | 105.4 | 6 | 4 | 10 | 7 | 10 | 4 | 0.2848 | 0.2624 | 0.548 | 0.443 |
| | 36 | 41.9 | 10 | 5 | 12 | 5 | 14 | 6 | 0.2858 | 0.2624 | 0.596 | 0.371 |
| | 48 | 33.3 | 16 | 8 | 15 | 6 | 17 | 8 | 0.2886 | 0.2623 | 0.656 | 0.349 |
| | 60 | 31.2 | 21 | 11 | 16 | 6 | 23 | 11 | 0.2896 | 0.2623 | 0.662 | 0.338 |

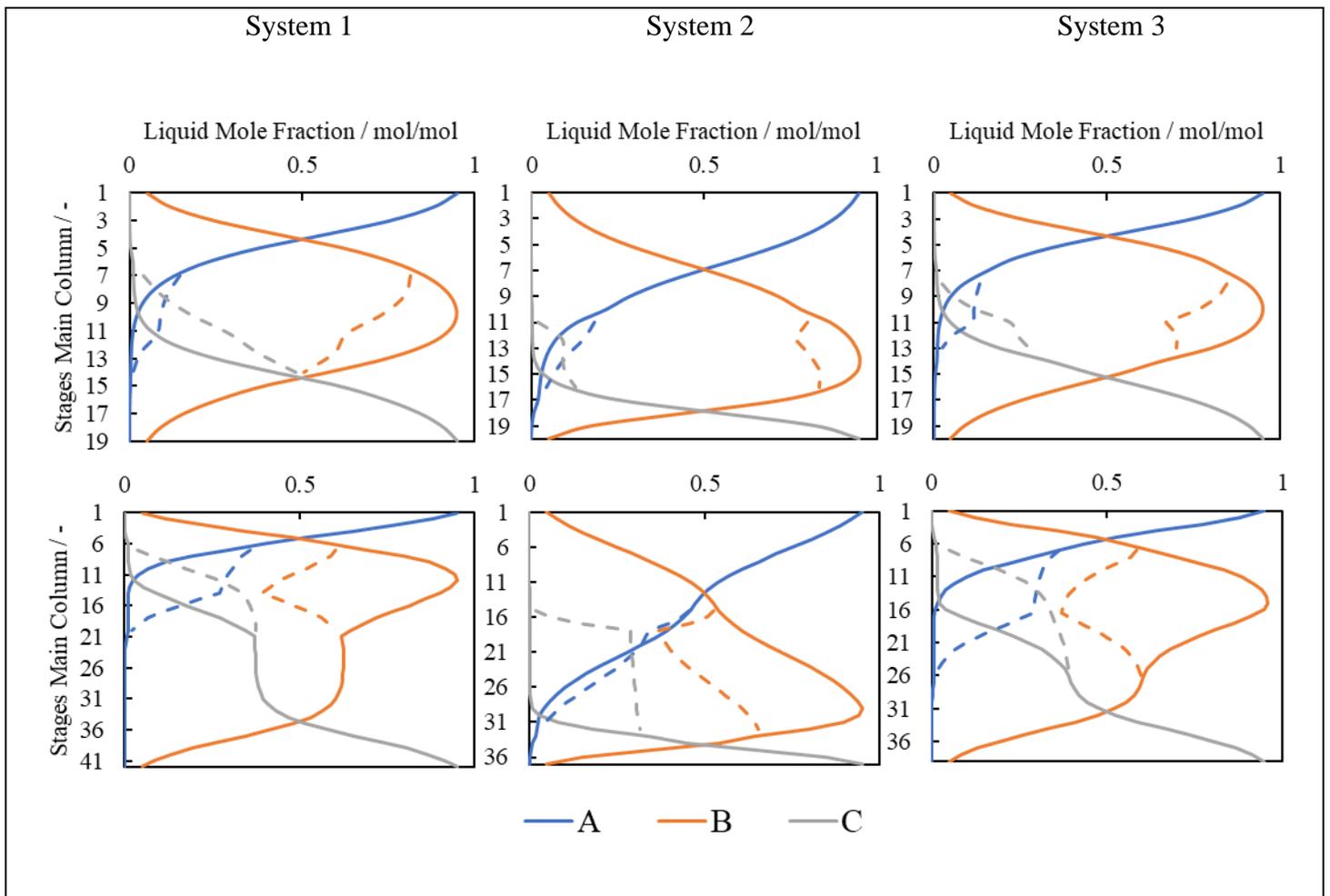

Figure 4: Nominal composition profiles. Solid lines: main column. Dashed lines: prefractionator.

The similarity between Systems 1 and 3 becomes obvious from the profiles: at low numbers of stages, the profiles are symmetrical, with a similar number of stages for AB and BC split. The concentration of A is low in the entire prefractionator, with around 20 mol% A and 80 mol% B at the upper end. There is a difference in the lower prefractionator range: in System 1, the concentration of C increases strongly, at the lower end there is an approximately equimolar mixture of B and C, while in System 3 the concentration of B remains clearly the highest. Towards the high number of stages, the stages assigned to the BC split in particular increase; in System 3 a pinch zone is indicated, in System 1 it is already pronounced.

For System 2, the difference in the AB and BC VLE is generally recognizable, and the high concentration of B in the entire prefractionator is also noticeable. At high stages, a beginning pinch zone is slightly visible for the AB split. It is also noticeable that the side product is only contaminated with A instead of light and heavy boiler as in the other two mixtures.

Particularly in the case of high stages, the BC separations generally tend to be carried out in the upper prefractionator sections with an almost constant A concentration, while the AB separations tends to be carried out in the lower section with a constant C concentration. This is again particularly pronounced in System 2, where hardly any stages are required for the easy BC separation above the feed stage in the prefractionator, but many are required below the feed stage. The heavy-boiler concentration is practically unchanged here.

### 2.4 Rigorous Process Simulation

All simulations of the optimized DWCs under uncertain thermodynamic models were performed in MATLAB® R2022a (MATLAB in the following). For this, the stage-to-stage scheme as explained in [24] and used in [9] for the multi-objective optimization was employed. For simulation purposes without optimization, the objectives (reboiler duty and stage allocation) are set constant to the value of the respective optimized DWC. Consequently, the stage allocation is fixed and all variables for solving the MESH equations are continuous. MATLAB's own fmincon solver was used in this case. As specifications, two product flow rates, the heat duty, the liquid split and the vapor split are fixed. For a better convergence of the simulations the vapor flowrates below the dividing wall and the liquid flowrates above the dividing wall were considered to be approximately constant, allowing the split to be calculated with the vapor leaving the reboiler and the liquid leaving the condenser, respectively.

### 2.5 Examined Uncertainties

Uncertainties in important caloric properties (i.e., liquid, vapor and evaporation enthalpy, although only the latter two are considered here, with the liquid enthalpy resulting from them) and in the vapor-liquid equilibrium were selected for the investigations. With regard to modified

Raoult's law for non-ideal liquid solutions at near-ambient pressure $p$, these are the pure substance vapor pressures $p_i°$ and the activity coefficients $\gamma_i$.

$$y_i \cdot p = \gamma_i \cdot x_i \cdot p_i° \tag{1}$$

For each model, realistic uncertainties (i.e., as they might actually be present between the properties as calculated by the respective model and reality) had to be determined first. For this, the differences between the values from the respective model in Aspen and available experimental data sets from DDB and NIST were evaluated, this is shown exemplarily in Fig. 5 for the binary VLE of System 1. While relatively constant percentual fluctuation ranges were found for pure substance vapor pressures and enthalpies over the entire temperature range, the uncertainty in the VLE is the highest at very low concentrations of a component. Since findings of a general nature are required (and the ranges determined for all three mixtures are within a similar range and also consistent with other mixtures from literature [16, 18]), the same uncertainties as shown in Tab. 3 are used for each mixture.

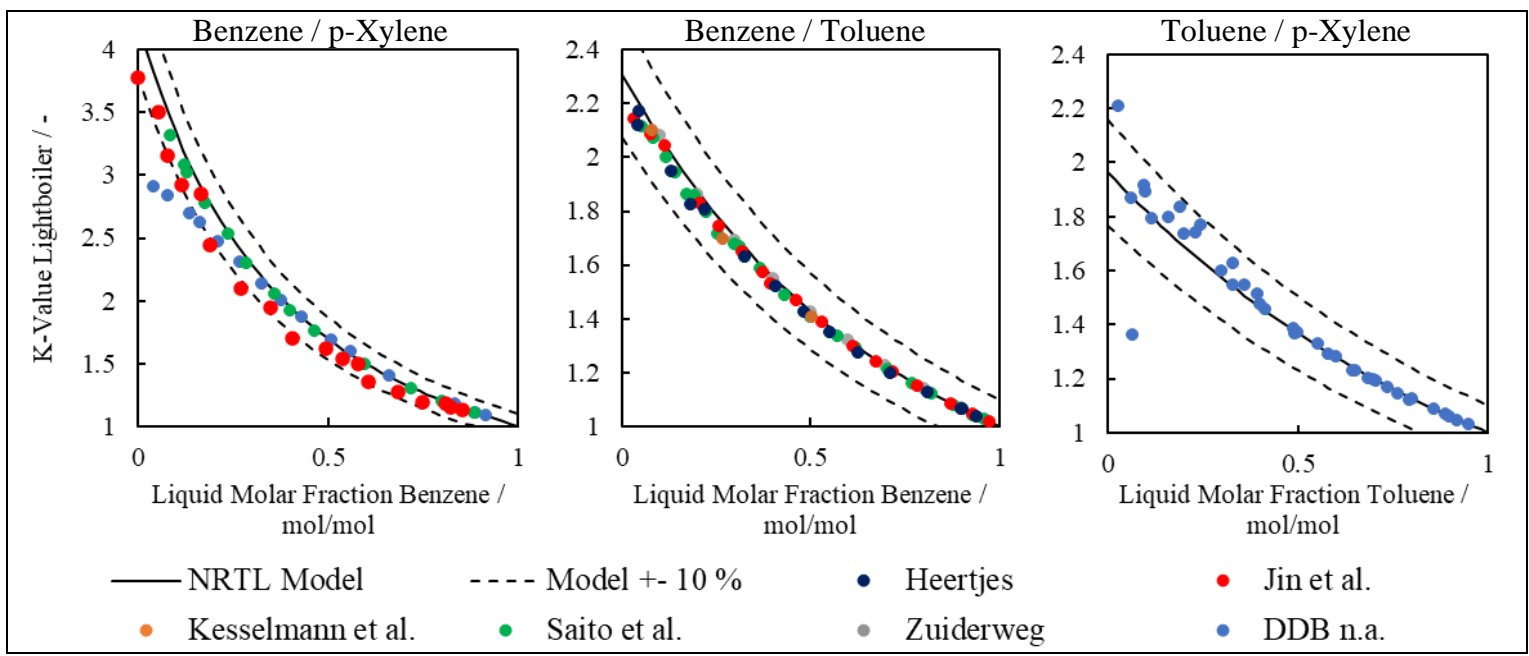

Figure 5: Determination of realistic uncertainties using the example of binary VLE of System 1. References: Heertjes [25], Jin et al. [26], Kesselmann et al. [27], Saito et al. [28], Zuiderweg [29]. DDB n.a. means data retrieved from DDB without author information.

Table 3: Thermodynamic properties and considered uncertainties.

| Variable | Uncertainty |
|---|---|
| Binary Infinite Dilution Activity Coefficient $\gamma_{ij}^\infty$ | ± 10 % |
| Pure Vapor Pressure $p_i^\circ$ | ± 2 % |
| Enthalpy of Vaporization $\Delta h_v$ | ± 1 % |
| Vapor Enthalpy $h_v$ | ± 2 % |

Uncertainties of pure component properties are added as a simple multiplication term to the property as calculated by the nominal model, i.e., the perturbed vapor pressure $p_i^{\circ,*}$ is varied between 0.98 and $1.02 \cdot p_i^{\circ,m}$ (nominal vapor pressure). Uncertain mixture properties on the other hand are treated as proposed by Burger et al. [18]. A perturbation term $\ln \gamma_i^p$ is added to the activity coefficient as calculated from the NRTL model $\ln \gamma_i^m$ according to Eq. 1, resulting in the perturbed activity coefficient $\ln \gamma_i^*$ used for the VLE calculation:

$$\ln \gamma_i^* = \ln \gamma_i^m + \ln \gamma_i^p \tag{2}$$

The perturbation term $\gamma_i^p$ is calculated using a Margules model, which takes into account a maximum deviation at $x_i \to 0$ and $\gamma_i = 1$ for pure substances as well as all three binary systems and ensures thermodynamic consistency, see Eq. 2:

$$\ln \gamma_i^p = \sum_{j=1}^{N} \sum_{k=1}^{N} x_j x_k \left[ 2 x_i A_{ji} + x_j (A_{ij} - A_{kj}) - x_K A_{jk} \right] \tag{3}$$

$N$ is the number of components and the factors $A$ can, due to the construction of the perturbation, directly be calculated from the desired perturbation in the binary infinite dilution activity coefficient $\gamma_{ij}^{p,\infty}$, i.e., where the largest perturbation is expected:

$$A_{ij} = \ln \gamma_{ij}^{p,\infty} \tag{4}$$

Since the maximum deviations for the case of infinite dilution should be approx. ± 10 % for all components and binary sub-mixtures as explained above, this leads to values for all $A_{ij}$ between $\ln(0.9) = -0.1$ and $\ln(1.1) = 0.1$.

### 2.6 Effect of Uncertain Pure Vapor Pressures and Activity Coefficients on VLE

As shown later in the results, the effects of uncertain caloric models proved to be much smaller than those of uncertainties directly concerning the VLE and were classified as insignificant and not used for combined scenarios. For this reason, only the two models with significant impact, the activity coefficients and vapor pressures, are discussed in more detail here. Supporting material provides associated diagrams. Since the effects of the activity coefficient perturbation term expressed by the Margules equation are not necessarily intuitive, Supporting Material 1 shows first the impact of positive and negative factors $A$ on the binary boundary systems. The correlations visible there are the same for all mixtures: negative $A$ result naturally in activity coefficients lower than in the nominal model, while positive $A$ lead to higher ones. For mixed factors, $A_{ij}$ predominates at lower concentrations of component i, while $A_{ji}$ predominates at lower concentrations of component j. The presentation in the ternary space has been omitted for reasons of clarity. The changes in the curves appear to be quite clear, but are low in absolute terms due to the ideality of the mixtures. Disturbed $T$-$x$-$y$-diagrams, also shown in Supporting Material 1, provide more information regarding the realistic nature of these uncertainties. The perturbed activity coefficients naturally lead to a mixture behaviour that is more non-ideal than calculated with the nominal model. $A_{ij}$ mainly influences the boiling line at low concentrations of i (up to approx. 50 mol% in the liquid phase), whereas $A_{ji}$ mainly influences the boiling line at corresponding vapor concentrations of j up to approx. 50 %. The influence of both factors is equally directed and is amplified accordingly with equally combined values: if both factors are negative, the boiling and dew lines shift slightly upwards; if both are positive, vice versa. If, on the other hand, the factors are inversely combined, the boiling line in particular is "dented", i.e., it is higher in one range and lower in another. This corresponds to a slightly more non-ideal behavior of the real mixture compared to the model and to the findings of Mathias [16].

*T-x-y*-diagrams for ± 2 % pure vapor pressure perturbation for the binary AC sub-system of System 1 can also be found in Supporting Material 1. The basic effects visible are analogous for the other binary boundary systems of the three mixtures. The perturbation $p_i^{\circ,*} = 0.98 \cdot p_i^{\circ,m}$ (i.e., the real vapor pressure of the substance is 2 % lower than predicted by the model) corresponds to an increase, $p_i^{\circ,*} = 1.02 \cdot p_i^{\circ,m}$ (vice versa) to a decrease of the real VLE in the direction of the corresponding pure substance compared to the nominal model. Opposite deviations correspond to a rotation, equal deviations to a shift of the entire VLE upwards or downwards. A higher real vapor pressure of the light boiler and a lower one of the heavy boiler compared to the model means an easier separation and vice versa to a more difficult one than predicted.

In general, the effects of the chosen maximum perturbation are barely visible for both models. It is therefore found that the imposed perturbations are suitable, i.e., do not lead to unrealistically changed characteristics. As an example, Fig. 6 shows the *T-x-y*-diagram for – 2 % perturbation in both vapor pressures in the sub-system Benzene-p-Xylene. The chosen uncertainties therefore represent cases in which the model could be assumed to be well suited for design or a range, in which a model error could be found. This should be kept in mind when considering the effects on the separation result.

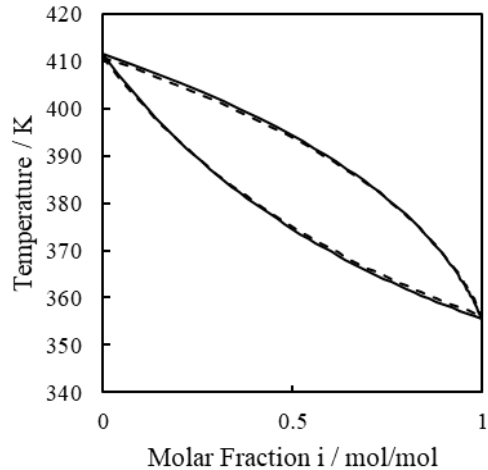

Figure 6: Impact of perturbed pure vapor pressures on the *T-x-y*-diagrams of the AC sub-system of System 1 (Benzene – p-Xylene). Solid lines: nominal model. Dashed lines: 2 % higher vapor pressure of C and 2 % lower for A.

## 2.7  Data Visualization: Self-Organizing Patch Plots

The results of combined scenarios are presented in Self-Organizing Patch Plots (SOPPs), a method that allows a data point with any number of dimensions to be represented. For the authors, this was found to be the best option for the presentation of so many data points, each with multiple dimensions. Network diagrams, for example, appeared to be incomprehensible. There is one diagram for each dimension. Each data point can be found as a single patch in each diagram, i.e., in the same place and in the same shape and size. The respective value is represented exclusively by the colour scale; the shape and size of the patches have no significance in this respect. For more details, readers are referred to Stöbener et al. [30], where the calculation was described, and to Ränger et al. [31], where a detailed explanation of the background and how SOPPs are to be read can be found.

## 3 Results and Discussion

### 3.1 Sensitivity to Individual Uncertainties

In the following, the resulting product purities from the steady-state simulations under manipulated thermodynamic models are discussed. First, individual influences (i.e., only one model is uncertain) were examined. For this, the same case studies were simulated for each of the four Pareto-optimal DWCs of each mixture. Tab. 4 shows the maximum losses per mixture and product $\Delta x_i = 0.95 - x_i$. Note that these results always represent the worst case of all possible combinations of the respective property of all three components. For example, in the case of the vapor pressure, all possible combinations of nominal and positively and negatively deviating $p_A°$, $p_B°$ and $p_C°$ have been screened. I.e., the 27 scenarios of possible combinations were formed and simulated successively. Similarly, all possible combinations of deviating activity coefficient were used (three possible values for six different factors $A$). It must also be mentioned that in some cases, of course, higher purities than the nominal result, since certain perturbations particularly in the vapor pressure and the VLE can also be advantageous for the separation. However, since the focus here is on operational risks due to unrecognized or underestimated uncertainties, only purity losses are discussed in the following. For the subsequent simulations of combined uncertainties, only the models causing more than 1 mol% loss in at least one case were chosen (marked bold in the table). In the range of uncertainties examined here, these are the properties influencing the VLE, but not the two caloric properties (enthalpy of vaporization and vapor enthalpy). Of course, no claim of general validity is made here. With other mixtures or different uncertainties, the selection of "relevant" models as well as the absolute values will deviate - however, the aim here is to identify relations and, above all, correlations to column and mixture characteristics. As in the previous work [9], a significant dependence of the purity losses of some products on the total stage number of the column was observed in some cases. This is the case, when perturbations affect the VLE (activity coefficients and vapor pressures) or *V/L* ratio inside the DWC (enthalpy of vaporization - a clear dependence on the number of stages was observed, but due to the low losses in absolute terms, the influence was nevertheless classified as insignificant). The dependencies are indicated in the table as follows: If a factor of at least 1.5 was found between the losses at high and low stages ($\Delta x_i^{highN}/ \Delta x_i^{lowN}$), this is indicated with +. A factor of at least 3 is indicated with ++, while no labeling means no significant dependence. The dependencies of losses due to uncertain activity coefficients and the pure vapor pressures on the total number of stages are shown in more detail in Fig. 7 (worst case combination is shown in each case).

Table 4: Maximum losses Δxi / mol% in product purities for uncertain thermodynamic properties. These result in case of high stage numbers; exceptions are indicated with *. Indicated are dependencies on the total stage number, where ++ means a factor of at least 3 between the losses at a high and those at low number of stages, + a factor of at least 1.5.

|  | S1 | | | S2 | | | S3 | | |
|---|---|---|---|---|---|---|---|---|---|
|  | A | B | C | A | B | C | A | B | C |
| **Activity Coefficient** | 1.0 + | 2.1 + | 1.2 ++ | 3.0 + | 3.2 + | 1.2 | 1.0 | 2.1 + | 1.8 ++ |
| **Pure Vapor Pressure** | 0.6 | 1.1 + | 1.0 + | 2.3 + | 2.4 + | 0.2* | 0.6 | 1.5 + | 1.4 + |
| Enthalpy of Vaporization | 0.1 | 0.3 ++ | 0.2 + | 0.4 + | 0.4 ++ | 0.1* | 0.1 | 0.5 ++ | 0.4 ++ |
| Vapor Enthalpy | <<1 | <<1 | <<1 | <<1 | <<1 | <<1 | <<1 | 0.1 | 0.1 |

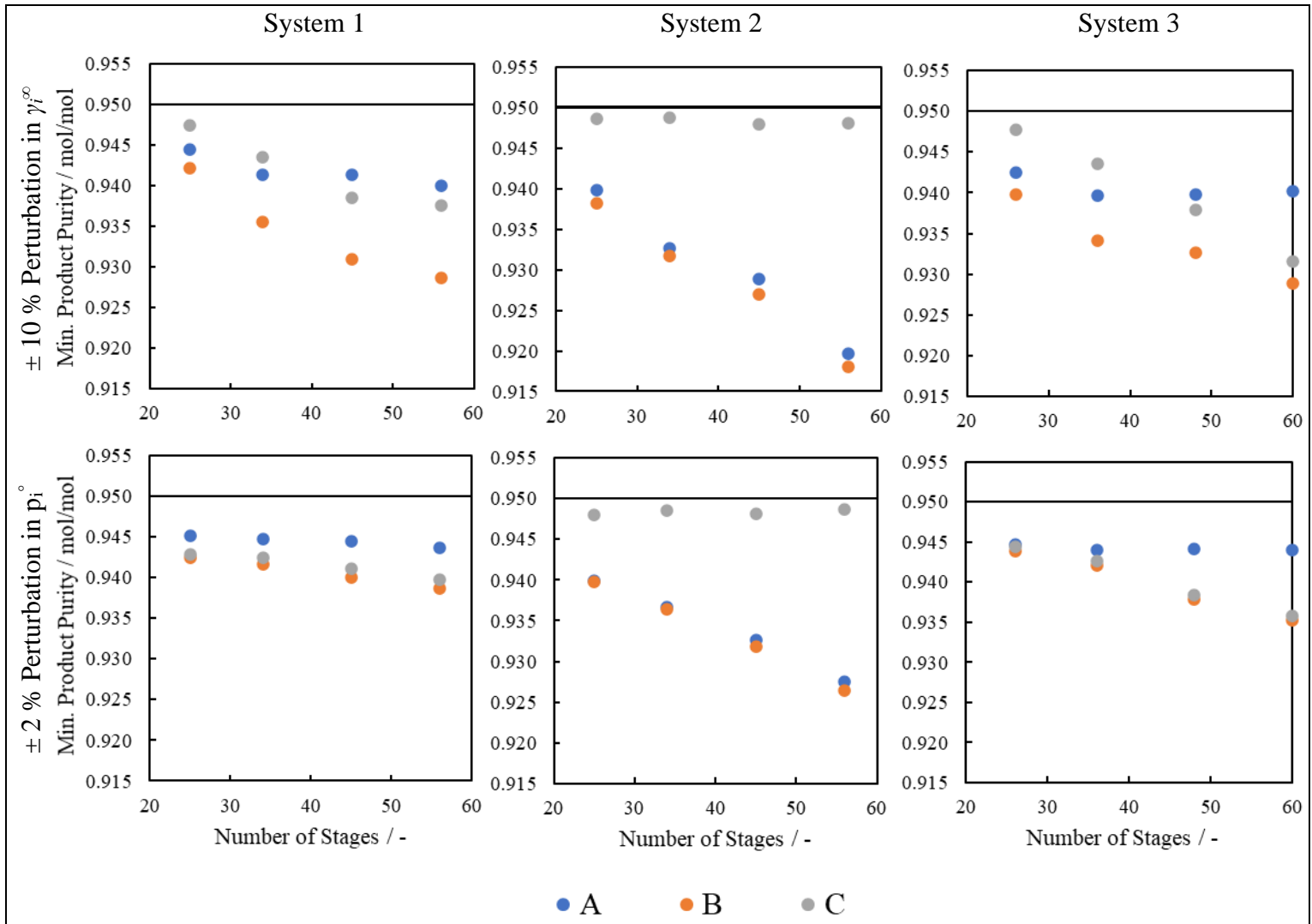

Figure 7: Maximum losses in product purities in case of individual uncertainties in thermodynamics with dependence on the total stage number. Black lines: nominal case.

The following can be stated regarding these overall worst-cases:

- Stage-dependent patterns:
    - The higher the number of stages, the higher the possible loss of purity.
- Mixture-dependent patterns:
    - The stage dependency mainly affects the split with the higher $V_{min}$ peak, i.e., two of the products.
    - Product C of System 2 is insensitive to any uncertainty in thermodynamics and resulting contaminations only affect products A and B. This has been similarly observed in [9] for deviations in some process variables and is again the result of

one binary main column separation being significantly easier than the other one: the difference between working and equilibrium line is larger in the associated section (i.e., the driving force per stage is higher) than in the other, difficult separation, which, however, determines the vapor requirement of the column and thus also at least roughly the vapor to liquid ratio of the easier separation. As a result, deviations from the nominal case cannot cause pinches, unlike in the case of the difficult separation, but in any case, the relative change in the position of the VLE and operation line is significantly lower. Even in the disturbed case, a high driving force per stage remains and the nominal purity can (almost) be achieved despite the deviation. An example is shown in the following section in Fig. 9.

- o Due to the narrow AB VLE, System 2 shows the most pronounced stage dependency and the highest absolute maximum losses per number of stages.
- o Systems 1 and 3 with similarly shaped $V_{min}$ diagrams and VLE show a very similar behavior.
- o In System 1 and 3 a particularity with regard to the activity coefficients can be seen: at low stage numbers, product C is more robust even though it is part of the split with the highest minimum energy demand. This changes for higher stage numbers.

- Product-dependent patterns:
  - o Product B has the highest maximum losses in each mixture and for each number of stages. In System 1 and 3 this is due to the fact that it can contain increased contaminations from both A and C, in System 2 the difficult AB separation is the reason.

A more detailed analysis of the correlations between specific combinations of perturbations and the different products of the mixtures was done, i.e., results were arranged to different subsets of vapor pressure or activity coefficient perturbations. The respective diagrams are provided with the Supporting Material 2. First, observations regarding differently deviating pure vapor pressures are summarized. Specific design-related and summarized findings can be found in section 3.2 as well as in the conclusion.

- Mixture-dependent patterns:
  - In the decomposed view of the diagrams in the Supporting Material 2, it is visible that the dependencies of System 1 and 3 are both quantitatively and qualitatively very similar.
  - In System 1 and 3, uncertain pure vapor pressures potentially affect all products. Regarding product B, an uncertain vapor pressure of substance C is particularly disadvantageous in both mixtures (BC is the split with highest $V_{min}$ peak). A nominal vapor pressure of substance B is relatively insignificant, but gains slightly in importance for higher stages, especially in System 3.
  - Product C of System 2 is not sensitive at all and impurities in product A and B result from the respective other one. Therefore, the order and effect of the different subsets is the same for product A and B, which is not the case for System 1 and 3. Here, the vapor pressure of substance C plays practically no role in the investigated range.
- Product-dependent patterns:
  - If the own and "adjacent" pure vapor pressure of component B is nominal, the nominal product purity is achieved for product A and C independent on the stage number. If both are uncertain, worst cases can result - i.e., product A is not dependent on $p_C°$, product C is not dependent on $p_A°$. All other possibilities (where either the own or the vapor pressure of B is uncertain) are in between and very similar in each case.
  - A negatively deviating (lower) $p_A°$ in combination with a positively deviating $p_B°$ is most disadvantageous for product A, for product C, a negatively deviating $p_B°$ in combination with a positively deviating $p_C°$. This is obvious given the impact of the vapor pressure in the VLE and the distinction between light and heavy key. For product B there is no mixture-independent pattern.

For different subsets of activity coefficients (diagrams also in the Supporting Material 2), the following observations can be summarized:

- Stage-dependent patterns:
    - For all columns with low numbers of stages, the factors $A_{AC}$ and $A_{CA}$ have no influence; the one of $A_{BC}$ and $A_{CB}$ is also very limited. The factors $A_{AB}$ and $A_{BA}$ have a significant influence, but only on product A and B – this leads to product C being in general most robust at low stages as observed in Fig. 7.
- Stage/Mixture-dependent patterns:
    - In System 1, mainly the influence of $A_{BC}$ and $A_{CB}$ (BC is the highest $V_{min}$ peak and by far the most absolute and relative stages are found in the associated column section) is increasing with the number of stages. This is similar but not as clear in System 3.
    - In System 2, the influence of $A_{AC}$ and $A_{CA}$ as well as $A_{BC}$ and $A_{CB}$ increases with increasing stages. The one of $A_{AB}$ and $A_{BA}$ (highest $V_{min}$ peak) reaches a minimum at 45 stages due to a phenomenon described below and then rises again.
    - In both Systems 1 and 3, the factors $A_{BC}$ and $A_{CB}$ can have a positive effect at high stage numbers on product B and C: for System 1, this is the case for both factors deviating negatively, for System 3 it is a negative $A_{BC}$ and a positive $A_{CB}$. Interestingly, in both cases these combinations are the (or among the) worst ones in case of low numbers of stages.
    - In System 2, this phenomenon can be observed inversely with the factors $A_{AB}$ and $A_{BA}$ (i.e., each time the factors belonging to the higher $V_{min}$ peak and the narrower VLE): at low stages, the combination of both factors deviating positively leads to higher than nominal purities of products A and B, while both factors deviating

negatively leads to the greatest loss. For the column with the highest number of stages, the opposite applies (higher than nominal purities are no longer achieved), i.e., the diagram is mirrored horizontally for higher numbers of stages.

It is assumed that this phenomenon is due to different compositions of the connecting streams between prefractionator and main column (which can be considered as binary feeds to the upper and lower main column) at low and high numbers of stages: Depending on this feed composition and the resulting operations in the section above and below the feed stage, a combination of modified activity coefficients can lead to a favorable change in the VLE in one case and a disadvantageous one in the other.

### 3.2 Influence of Simultaneously Disturbed Parameters

Simultaneous uncertainties in activity coefficients and pure vapor pressures were simulated next in all possible combinations. Different to the previous section, nominal values were no longer calculated for combined scenarios due to the resulting number of simulations (512 instead of 19 683). Fig. 8 shows the worst-case result per product for all three mixtures for these combined uncertainties.

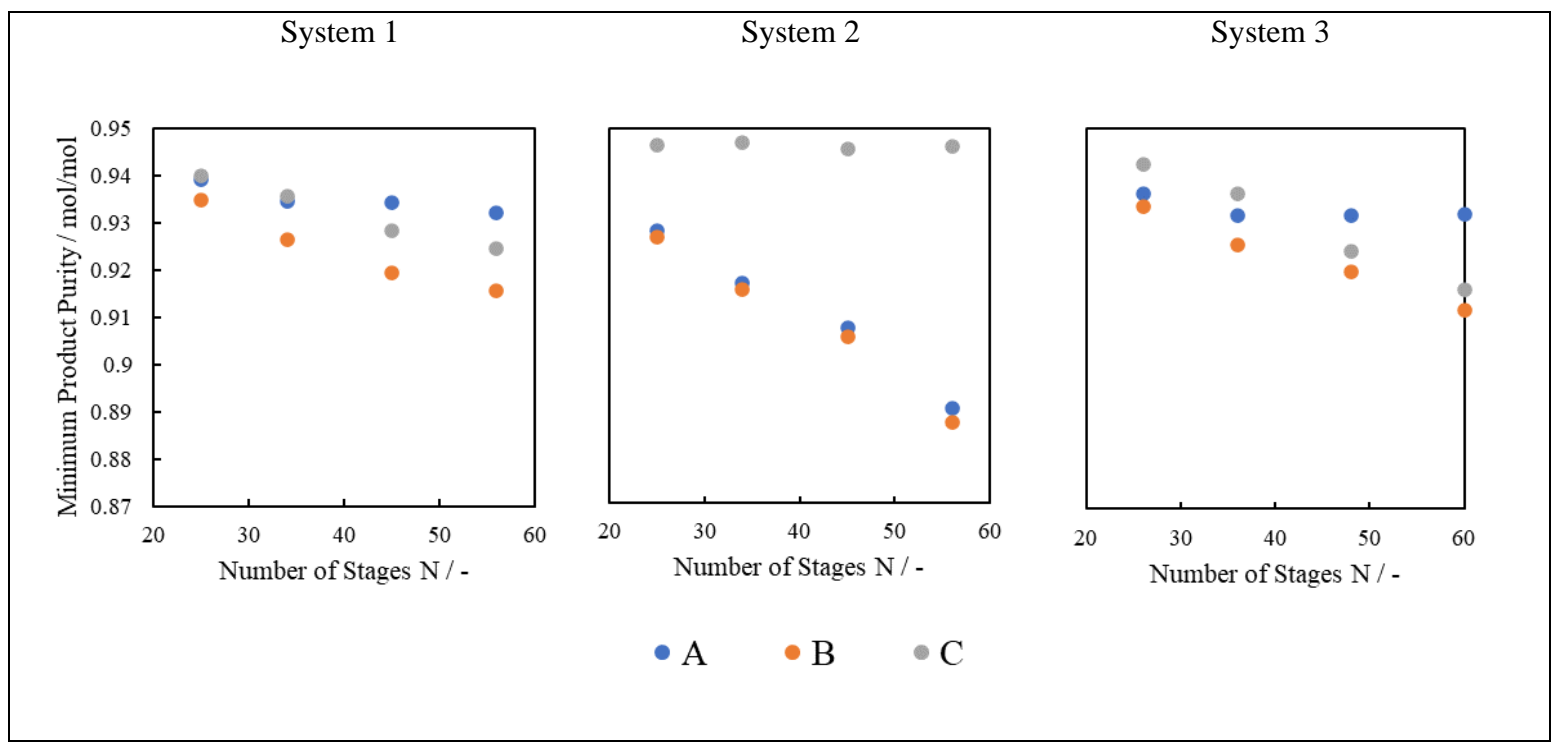

Figure 8: Maximum influence of combined uncertainties in thermodynamic properties.

As in the previous work investigating the influence of deviating process variables [9], the influences of the combined uncertainties correspond roughly to an addition of individual influences. Here, a closer look is taken at the robustness of product C of System 2, where even combined uncertainties lead to purity losses of < 0.5 mol%, regardless of the number of stages. It was first investigated which deviations are generally disadvantageous for product C, i.e., which lead to the worst case for each mixture and number of stages. The overlap of all mixtures consists of an increased vapor pressure of C, a reduced one of B and a factor $A_{BC} = 0.1$. This combination was simulated using the columns with high stage numbers of all three mixtures. Fig. 9 shows the resulting McCabe-Thiele diagrams (BC VLE) of the lower main column section of System 1 (System 3 analogous) and System 2. The data points show the concentration profiles read out from simulations. The representation in McCabe-Thiele diagrams is possible, because in the column sections shown, almost binary separations take place - above the side product, there is almost no component C and below it almost no A. Nevertheless, the traces of the third component were excluded, i.e., the fractions of the two main components together were normalized to 100 %. The deviations used lead in System 1 to a loss of several stages at the feed point due to an already existing pinch zone; due to the almost parallel VLE, the resulting shift remains until product C is discharged. In System 2, on the other hand, a high driving force remains when applying the same relative deviations. Increased B concentrations on the second and third stages from the bottom are practically compensated when reaching the bottom due to the very high VLE gradient in this concentration range. However, in the two mixtures where AB is the easier separation, the difference to the BC VLE is only marginal and the driving forces present in upper and lower main column in the nominal case are similar. As a result, System 1 and 3 do not provide a similarly robust product A.

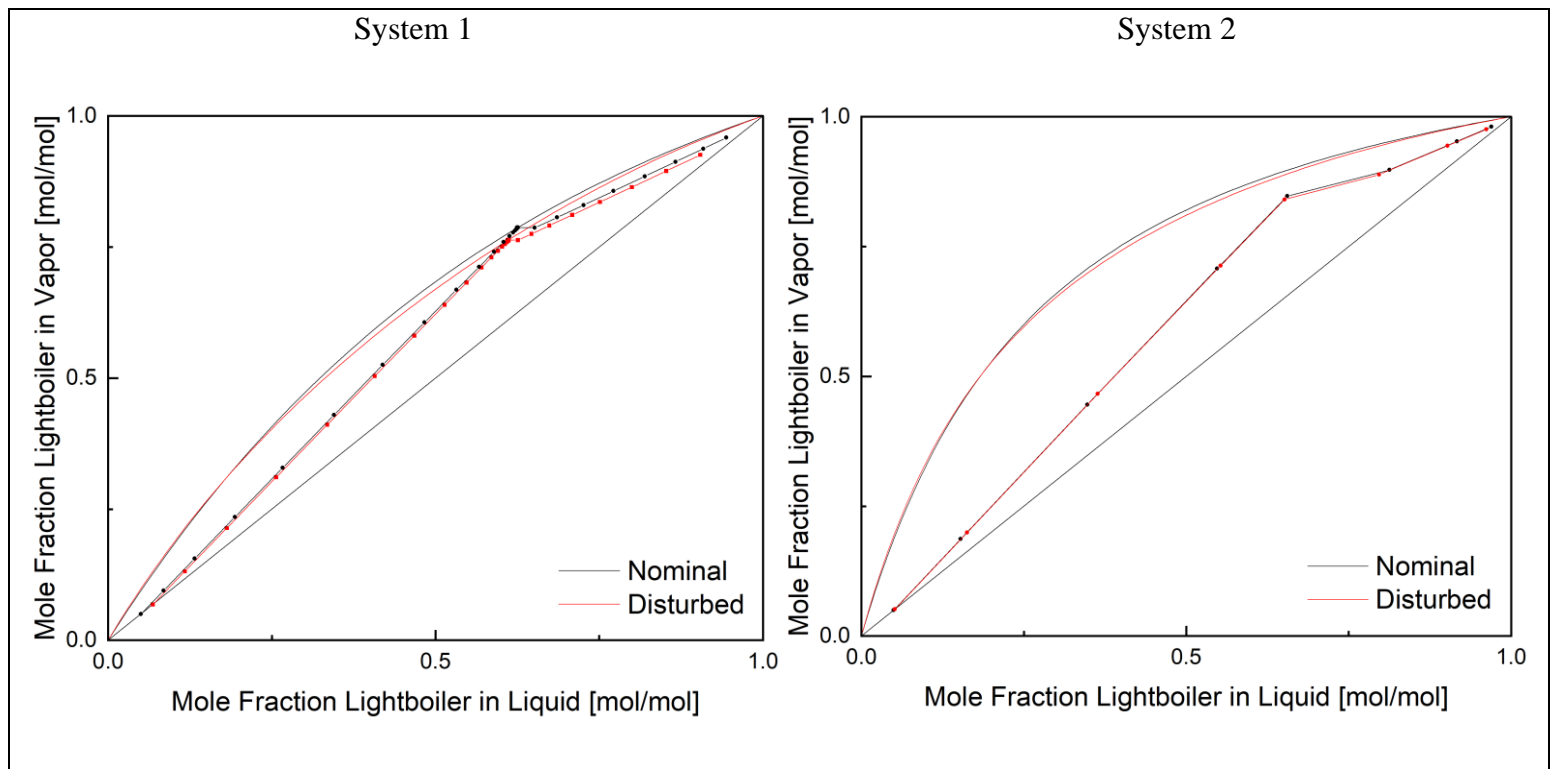

Figure 9: VLE of BC separation and operation lines in lower main column sections of System 1 and 2. Shown are the nominal situations and the "common worst case" for product C: higher $p_C°$, lower $p_B°$ and $A_{23} = 0.1$.

The representation of all calculated scenarios and their effects is shown in Fig. 10 - 11 using SOPPs. Fig. 10 shows the calculated scenarios, i.e., the input values for the simulations. Note that there are only values of 0 and 1, since only the maximum positive (corresponding to 1) and negative (corresponding to 0) deviations were used in each case, but neither values in between nor the nominal case. The scenarios can be most clearly divided into an upper half with high vapor pressure of B $p_B°$ and a lower half with low vapor pressure, as well as a right half with low $A_{AB}$ and a left half with high $A_{AB}$. The other distributions are complex and difficult to summarize. Regarding the resulting product purities in Fig. 11, the following can be summarized:

- Stage/Mixture-dependent patterns:
  - At low stages, the greatest risk is posed by an uncertain vapor pressure of the mid boiler $p_B°$ in combination with
    - an uncertain $A_{BA}$ (i.e., the mixture properties especially at low concentrations of B) where AB is the highest peak in the $V_{min}$ diagram and

- an uncertain $p_C°$ and $A_{BC}$ (i.e., the mixture properties especially at low concentrations of B) where BC is the highest peak.
  - At higher total numbers of stages (approx. $2.8 \cdot N_{min}$), the worst cases were found for unfavorable deviations in $p_B°$ in combination with
    - an unfavorably deviating $p_A°$, but an uncertain $A_{BA}$ having now the decisive influence in case of the mixture with AB as highest $V_{min}$ peak and
    - an unfavorably deviating $p_C°$, but an uncertain $A_{BC}$ now having the decisive influence in case of the mixtures with BC as the highest $V_{min}$ peak.

Some scenarios result in a higher purity at high than at low numbers of stages. Here, the increasingly positive (or at least less negative) effects of some model uncertainties visible in Fig. 2 – 4 in the Supporting Material 2 are present. However, these are not discussed further, as the focus is on operating risks on the one hand and uncertainties are not applicable instruments on the other.

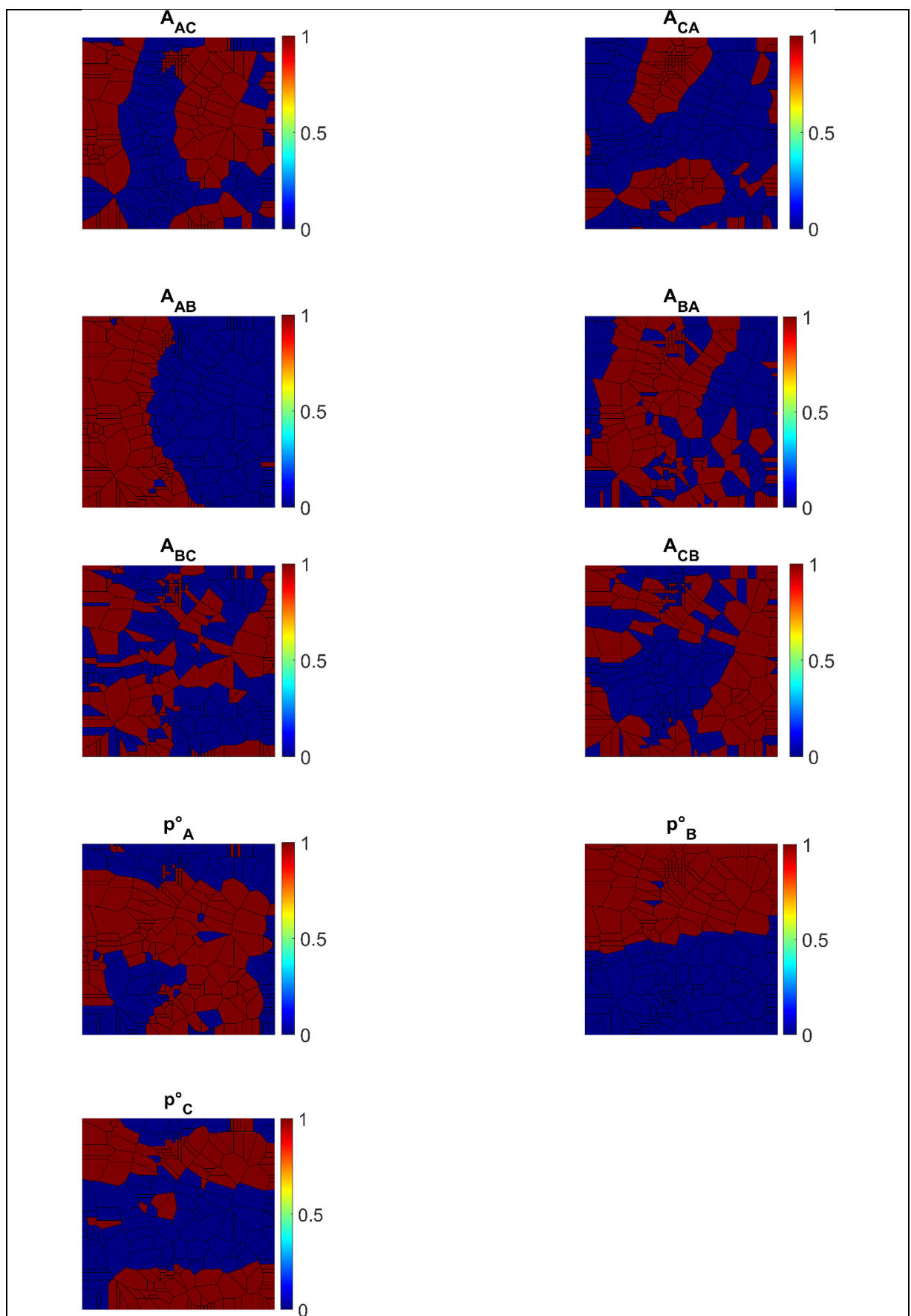

Figure 10: Scenarios of combined uncertainties in thermodynamic properties, results are shown in Figure 11.

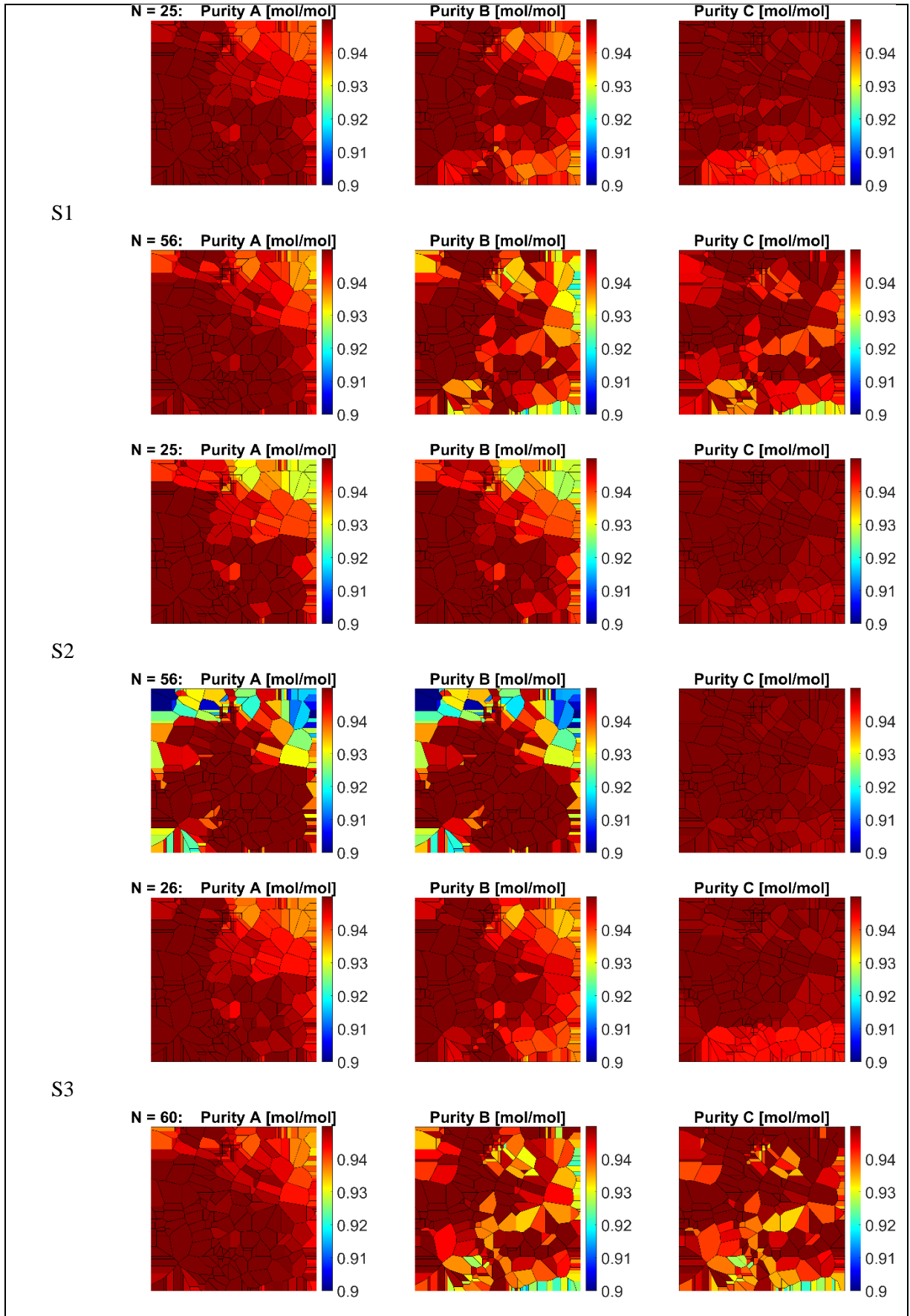

Figure 11: Effect of combined uncertainties in pure vapor pressures and activity coefficients on System 1 - 3, scenarios are shown in Figure 10.

## 4 Summary and Conclusion

This work is the second part of a research investigating relations between mixture properties, associated optimized Dividing Wall Columns and their sensitivity to deviations from the nominal design case. While the first part [9] introduced the mixtures and DWCs used for the simulations and presented results for deviations in process variables, the present work treats uncertainties in thermodynamic models. Therefore, the effects of possible unknown or accepted deviations between real substance or mixture behavior and thermodynamic models used for the design on the performance of the same Pareto-optimal DWCs as in [9] were investigated. These are four columns with different total number of stages for each of the three mixtures. Note that only ideal mixtures and steady-state simulations are used both in [9] and here. The extension to real mixtures and verifications with dynamic simulations are in progress and would have exceeded the scope of these works, which are intended to provide a framework.

Literature information, according to which models that have an influence on mixture properties in particular have a significant influence, is confirmed. The caloric properties examined - enthalpy of vapor and enthalpy of evaporation - were significantly less relevant in the chosen ranges of uncertainties. Regarding vapor pressures and liquid activity coefficients on the other hand, deviations that are hardly visible in *T-x-y*-diagrams caused purity losses of up to over 6 mol%, depending on the mixture and number of stages. The influence of the uncertainties was found to be dependent on the total number of stages (i.e., increasing with it) if the vapor-liquid equilibrium (vapor pressures and activity coefficients) or the vapor to liquid ratio inside the column (enthalpy of vaporization) is affected. In the previous work [9], the same dependency for manipulated vapor to liquid ratios was observed. Simplified, this stage dependency is therefore present when the position of equilibrium and operating lines are influenced in relation to each other. Consequently, a separation and the associated products are particularly affected, when the nominal operation line is already very close to the VLE. Since the vapor to liquid ratios in the different column sections are linked, the differences in the resulting impurities in the products are larger if one of the separations AB and BC is significantly easier than the other.

Also in accordance with the previous work, a significant influence of "the position of highest peak in the $V_{min}$ diagram", i.e., whether the light/medium boiler (AB) or the medium/heavy boiler (BC) separation has the higher minimum energy requirement, was found. The products belonging to this split are mainly affected by uncertain pure vapor pressures in general and by uncertain activity coefficients at least at higher numbers of stages. In principle, the top product is not influenced by the vapor pressure (within the range investigated) of the heavy boiler $p_C°$, and the bottom product is not influenced by that of the light boiler $p_A°$. The influence of the activity coefficients is complex with phenomena like one and the same combination of deviations having a positive influence at low stage numbers and a negative one at high stages or vice versa. The reason for this is assumed to be that the connecting streams between prefractionator and main column and accordingly the concentration ranges of the separations in the different DWC sections differ between low and high numbers of stages. Manipulated activity coefficients can lead to VLEs becoming wider in certain concentrations ranges and narrower in others – especially in the ternary space, a complex "curvy" behavior can result. Depending on the composition range covered in a certain column section, one and the same combination can be advantageous (increasing the distance between VLE and operation) or disadvantageous (decreasing the distance or leading even to a pinch).

Combined uncertainties in vapor pressures and activity coefficients showed in principle a superposition of the individual effects. It can be summarized that the influence of the investigated uncertainties at low numbers of stages (approx. 20 % above the minimum) was moderate and even combined uncertainties hardly led to purity losses of more than 1 mol%. At higher total numbers of stages (approx. 2.8 · $N_{min}$), 3.5 to over 6 mol% loss of purity was found, depending on the mixture.

The greatest risk of high losses comes from inaccuracies in the more difficult binary VLE (and the highest peak in the original $V_{min}$ diagram). Particular attention must be paid to mixture properties at rather low mid-boiler concentrations.

After considering the influences of process variables in [9] and thermodynamics, it is obvious that generalized statements - especially regarding design rules for an optimal compromise between approximation of the energetic optimum and operational flexibility - still require considerable research. Future work should focus on the following aspects:

- The analysis of the compensation possibility for the relevant disturbances by typical control variables without increasing the energy input or the (additional) increase in reboiler duty required to achieve the product purities - if possible at all.

- Verification of all findings obtained up to this point for other, non-ideal mixtures as well as in dynamic models.

- Examination of the possibility of different stage allocations inside the columns in combination with different internal splits leading to (almost) identical $N$-$Q$-combinations, but to a significantly different flexibility and/or adjustment potential. And, if this is the case, how an optimization problem would have to be formulated in order to find the superior solution in this respect.

# 5 Abbreviations and Symbols

Table 5: List of abbreviations

| Abbreviation | Description |
|---|---|
| CAPEX | Capital expenditures |
| (m)DWC | (multiple) Dividing Wall Column |
| LS | Liquid Split |
| MESH | Material balance, equilibrium and summation condition, heat balance |
| NRTL | Non-Random Two Liquid |
| OPEX | Operational expenditures |
| SOPP | Self-Organizing Patch Plot |
| VLE | Vapor-Liquid-Equilibrium |
| VS | Vapor Split |

Table 6: List of symbols

| Symbol | Description | Unit |
|---|---|---|
| $\alpha$ | Relative volatility | - |
| $A_{ij}$ | Margules coefficients for activity coefficient perturbation | - |
| $\gamma_i$ | Activity coefficients | - |
| $h_v$ | Vapor enthalpy | J/mol |
| $\Delta h_v$ | Enthalpy of vaporization | J/mol |
| $F$ | Feed flow | mol/s |
| $N$ | Number of theoretical stages | - |
| $p$ | Pressure | bar |
| $p_i^\circ$ | Pure substance vapor pressure | bar |
| $Q$ | Heat duty | kW |
| $R$ | Reflux ratio | - |
| $T$ | Temperature | K |
| $V$ | Vapor flow | kmol/h, mol/s |
| $x_i$ | Molar fraction in vapor phase | mol/mol |
| $y_i$ | Molar fraction in liquid phase | mol/mol |

Declaration of competing interests

The authors declare that they have no known competing financial interests or personal relationships that could have appeared to influence the work reported in this paper.

## 6 Author Contribution

**Lea Trescher:** Methodology, Software, Investigation, Data Curation, Writing – Original Draft, Writing – Review & Editing, Visualization. **David Mogalle:** Methodology, Software, Writing – Review & Editing. **Patrick Otto Ludl:** Methodology, Software, Writing – Review & Editing. **Tobias Seidel:** Conceptualization, Methodology, Software, Writing – Review & Editing. **Michael Bortz:** Conceptualization, Writing - Review & Editing, Project administration, Funding acquisition. **Thomas Grützner:** Conceptualization, Resources, Writing - Review & Editing, Supervision, Project administration, Funding acquisition.

## 7 Acknowledgements

We gratefully acknowledge the funding by Deutsche Forschungsgemeinschaft (DFG), project number 440334941.